\documentclass[titlepage]{article}
\usepackage[]{graphicx}
\usepackage[]{color}

\usepackage{amsmath,amsfonts,amssymb,amsthm,epsfig,epstopdf,titling,url,array}


\usepackage[section]{placeins}
\usepackage{setspace}
\theoremstyle{plain}
\newtheorem{thm}{Theorem}[section]
\newtheorem{lem}[thm]{Lemma}

\theoremstyle{definition}
\newtheorem{defn}{Definition}[section]

\theoremstyle{remark}

\makeatletter
\def\maxwidth{ %
  \ifdim\Gin@nat@width>\linewidth
    \linewidth
  \else
    \Gin@nat@width
  \fi
}
\makeatother

\definecolor{fgcolor}{rgb}{0.345, 0.345, 0.345}

\usepackage{framed}
\makeatletter
 {\par\unskip\endMakeFramed%
 \at@end@of@kframe}
\makeatother

\definecolor{shadecolor}{rgb}{.97, .97, .97}
\definecolor{messagecolor}{rgb}{0, 0, 0}
\definecolor{warningcolor}{rgb}{1, 0, 1}
\definecolor{errorcolor}{rgb}{1, 0, 0}

\usepackage{alltt}
\usepackage{enumitem}
\usepackage{amsthm}
\usepackage{amsmath}
\usepackage{amssymb}
\usepackage{amsfonts}

\usepackage[style=authoryear,maxcitenames=2, doi=false]{biblatex}

\bibliography{BlipVar}

\usepackage[letterpaper, portrait, lmargin=1in, rmargin=1in,
bmargin = 1.35in, tmargin = 1.35in]{geometry}
\usepackage[english]{babel}
\usepackage{graphicx}
\usepackage{float}

\usepackage{caption}

\IfFileExists{upquote.sty}{\usepackage{upquote}}{}

\title{An Easy Implementation of CV-TMLE}
\author{Jonathan Levy}

\begin{document}

\begin{titlepage}

\maketitle
\begin{abstract}
In the world of targeted learning, cross-validated targeted maximum likelihood estimators, CV-TMLE \parencite{Zheng:2010aa}, has a distinct advantage over TMLE \parencite{Laan:2006aa} in that one less condition is required of CV-TMLE in order to achieve asymptotic efficiency in the nonparametric or semiparametric settings.  CV-TMLE as originally formulated, consists of averaging usually 10 (for 10-fold cross-validation) parameter estimates, each of which is performed on a validation set separate from where the initial fit was trained.  The targeting step is usually performed as a pooled regression over all validation folds but in each fold, we separately evaluate any means as well as the parameter estimate.  One nice thing about CV-TMLE, is that we average 10 plug-in estimates so the plug-in quality of preserving the natural parameter bounds is respected. Our adjustment of this procedure also preserves the plug-in characteristic as well as avoids the donsker condtion.  The advantage of our procedure is the implementation of the targeting is identical to that of a regular TMLE, once all the validation set initial predictions have been formed.  In short, we stack the validation set predictions and pretend as if we have a regular TMLE, which is not necessarily quite a plug-in estimator on each fold but overall will perform asymptotically the same and might have some slight advantage,  a subject for future research.  In the case of average treatment effect, treatment specific mean and mean outcome under a stochastic intervention, the procedure overlaps exactly with the originally formulated CV-TMLE with a pooled regression for the targeting.  

\end{abstract}
\end{titlepage}

\newpage
\section*{Introduction}
The original formulation and theoretical results of cross-validated targeted maximum likelihood estimators, CV-TMLE \parencite{Zheng:2010aa}, leads to an algorithm for the CV-TMLE that generally requires 10 targeting steps for each of 10 validation folds for each iteration in an iterative targeted maximum likelihood estimators or TMLE \parencite{Laan:2006aa}.  Such can be done in one regression, which solves the efficient influence curve equation averaged over the validation folds.  However, in this pooled regression, we must keep track of the means used in each fold, making the process different than a regular TMLE, once the initial predictions have been formed.  The formulation of the CV-TMLE here-in leads to a simpler implementation of the targeting step in that the targeting step can be applied identically as for a regular TMLE once the initial estimates for each validation fold have been computed. The CV-TMLE as discussed here is currently implemented in the R software package of tlverse \parencite{sl3}. \\ 

\section{CV-TMLE Definition for General Estimation Problem}
We refer the reader to the following sources \parencite{Laan:2015aa,Laan:2015ab, Laan:2006aa, Laan:2011aa} for a more detailed look at the theory of TMLE and Zheng and van der Laan, 2010 for theory regarding CV-TMLE.  We consider iid data of the form $O\sim P \in \mathcal{M}$, nonparametric or semiparametric model and parameter mapping 

\[
\Psi(Q(\cdot)):\mathcal{M}\longrightarrow \mathbb{R}^d 
\]

Where $Q(P)$ is a model upon which the parameter depends.  If we consider $O=(W,A,Y)$ with outcome, $Y$, and treatment and covariates, $A$ and $W$, then the outcome model $\bar{Q}(A,W) = E_P[Y \mid A, W]$ and distribution of $W$, $Q_W$, would define $Q(P)$. We consider the canonical least favorable submodel \parencite{clfm} of model estimate $\hat{Q}(P_n)$ defined with one-dimensional $\epsilon$:

\[
\frac{d}{d\epsilon}L\left(\hat{Q}(P_n)(\epsilon)\right)\biggr\vert_{\epsilon=0} = \Vert D^*\left(\hat{Q}(P_n), \hat{g}(P_n)\right) \Vert_2
\]

This definition coincides with the least favorable submodel if the $d=1$ because in that case we will have
\[
\langle \frac{d}{d\epsilon}L\left(\hat{Q}(P_n)(\epsilon)\right)\biggr\vert_{\epsilon=0} \rangle  \supset \langle D^*\left(\hat{Q}(P_n), \hat{g}(P_n)\right) \rangle
\]

where the above $\Vert \cdot \Vert_2$ is the euclidean norm.  We then define a mapping $B_n \in {0,1}^n$ to be a random split of ${1,..,n}$.  The training set is defined as $\mathcal{T}=\{i : B_n(i)=0\}$ and the validation set, $\mathcal{V}=\{i : B_n(i)=1\}$.  As in Zheng 2010, $P_{n,B_n}^0$ and $P_{n,B_n}^1$ and the empirical distributions over $\mathcal{T}$ and $\mathcal{V}$ respectively. \\

The CV-TMLE estimator as in Zheng and van der Laan, 2010 is defined as
\[
\Psi^{k_n}(P_n) = E_{B_n} \Psi \left(\hat{Q}(P_{n,B_n}^0)(\overset{\rightarrow}{\epsilon_n}^{k_n})\right)
\]

where $\Psi  \left(\hat{Q}(P_{n,B_n}^0)(\overset{\rightarrow}{\epsilon_n}^{k_n})\right)$ is the plug-in estimator (usually an average of the plugged-in model over the validation set).  $\overset{\rightarrow}{\epsilon_n}^{k_n}$ denotes the kth iteration of fluctuation parameters, where $k$ could always be 1 if we use the one-step TMLE \parencite{Laan:2015ab}.  

\section{Illustrative Example}
We will now go through the CV-TMLE algorithm for the VTE, variance of treatment effect \parencite{blipvar}  Here, we notice that we never target the distribution of $W$, but rather use the unbiased estimator, the empirical distribution.  This is discussed in Zheng and van der Laan, 2010 so refer the reader there for more detail as to why this is often the case.  In short, the component of the efficient influence curve in the tangent space of mean 0 functions of $W$ \parencite{Vaart:2000aa} is given by $D^*_W(P) =  (b(P)(W) - E_P b(P)(W))^2$ where $b(P)(W) = E_P[Y \mid A = 1, W] - E_P[Y \mid A = 0, W]$.  For any approximation to this function, its empirical mean will automatically be zero. We denote the following to avoid heavy notation:
\[
\bar{Q}^k_{B_n} = \hat{Q}(P_{n,B_n}^0)(\overset{\rightarrow}{\epsilon_n}^{k})
\]
is the approximation of the outcome model at the kth iteration.  This fit is entirely dependent on the training set $P_{n,B_n}^0$ observations and the fluctuations to the model, performed on the corresponding validation set. 
\[
\bar{Q}^k_{1,B_n} = \hat{Q}(P_n)(\overset{\rightarrow}{\epsilon_n}^{k})
\]
is the approximation of the outcome model at the kth iteration.  We will see it actually depends on $P_{n,B_n}^0$ and $P_{n,B_n}^1 \hat{B}(P_{n,B_n}^0)(\overset{\rightarrow}{\epsilon_n}^{k})$, and hence the entire empirical draw of the data. 

\begin{eqnarray*}
\hat{b}^k_{B_n}(W) &=& \bar{Q}^k_{B_n}(1,W) - \bar{Q}^k_{B_n}(0,W)\\
\hat{b}^k_{1,B_n}(W) &=& \bar{Q}^k_{1,B_n}(1,W) - \bar{Q}^k_{1,B_n}(0,W)\\
\hat{g}_{B_n}(A \mid W) &=& \hat{g}(P_{n,B_n}^0)(A \mid W) 
\end{eqnarray*}

\begin{itemize}
\item
STEP 1: Initial estimates

For each split, $B_{n}$ as in standard 10-fold cross-validation, we use an ensemble learning package such as sl3 \parencite{sl3} or SuperLearner \parencite{Eric-Polley:2017aa} to fit a model on the training set, denoting the model as $P_{n, B_{n}}^0$.  In this case we will fit relevant factors of the likelihood, such as the propensity score and outcome model, but not the distribution of covariates, $W$.  For those, we use the empirical distribution as an unbiased estimator and will not target it.   Denote the initial fit of the $E_P[Y \mid A, W]$, which we denote $\bar{Q}^0_{B_n}$.  For both procedures the initial fits are all the same.  

\item
STEP 2: Check Tolerance

For each fold evaluate the so-called clever covariate: 

$H_{B_n}^k(A,W) = 2(\hat{b}^k_{B_n}(W) - P_{n,B_n}^1 \hat{b}^k_{B_n}) \frac{2A - 1}{\hat{g}_{B_n}(A \mid W)}$

and the influence curve approximation

 \[
D^*_{k, B_n}(O) = H_{B_n}^k(A,W)(Y -\bar{Q}^k_{B_n}(A,W))
\]

Our proposed procedure would do

$H_{1,B_n}^k(A,W) = 2(\hat{b}^k_{1,B_n}(W) - E_{B_n}P_{n,B_n}^1 \hat{b}^k_{1,B_n}) \frac{2A - 1}{\hat{g}_{B_n}(A \mid W)} $

and the alternate influence curve approximation

 \[
D_{k, B_n}(O)= H_{1,B_n}^k(A,W)(Y -\bar{Q}^k_{1,B_n}(A,W))
\]

Thus, in our procedure we need not keep track of the folds since the average within the clever covariate is merely taken over the entire sample.  Thus the process is identical to a TMLE once the initial estimates are made.  We just stack them on top of eachother and act is if it is all one initial fit as with the regular TMLE. \\

We then compute the influence curve approximation for each fold and take the sample mean.  Since the $T_W$ component, as stated above always has empirical average 0, we only need to take the mean of the component of the influence curve approximation in the tangent space, $T_Y =$ mean 0 functions of $Y \mid A, W$, which have finite variance \parencite{Vaart:2000aa}.  We then check if the mean of the influence curve is below the tolerance level, $\hat{\sigma}/n$ where $\hat{\sigma}$ is the sample standard deviation of the above influence curve computations.  This assures we stop the process when the bias is second order as any more fluctuations beyond that point are not helpful.  If we are below the tolerance we go to step 4.  Otherwise we continue onward.  
 
 \item
STEP 3: Targeting Step: Run a pooled logistic regression over all the folds with model: 

\[
Y = expit(logit \left(\bar{Q}^k_{B_n}(A,W) + \epsilon_n^k H(\bar{Q}^k_{B_n}(A \mid W) \right)
\]  

That is, a model which suppresses the intercept and uses and the initial predictions as the offset.  This is identical to our method, except we would use the slightly different clever covariate as stated above.  

Update all the predictions to form $\bar{Q}^{k+1}_{B_n}(A,W)$ or, as with our method $\bar{Q}^{k+1}_{1,B_n}(A,W)$.  

\item
STEP 4: Compute the estimate and CI:

\[
\Psi^{k_n}(P_n) = E_{B_n} \Psi \left(\hat{Q}(P_{n,B_n}^0)(\overset{\rightarrow}{\epsilon_n}^{k_n})\right)
\]

and estimate the standard error via the standard deviation of the influence curve in step 3 divided by root n, which we will just call $\hat{\sigma}/\sqrt{n}$ and form the confidence bands

\[
\Psi^{k_n}(P_n) \pm z_\alpha \hat{\sigma}/\sqrt{n}
\]

where $z_\alpha$ is the $1-\alpha/2$ normal quantile.  This entails computing the parameter separately per validation set before averaging the 10 estimates, i.e., compute the sample variance over the validation set for  $\hat{b}^k_{B_n}$, getting 10 estimates and then average them.  In our procedure we just have a list of n values of $\hat{b}^k_{1,B_n}$ and compute the sample variance over the entire sample.  

\end{itemize}

Thus we can see our procedure simplifies the targeting and, like the original formulation, solves the efficient influence curve equation, i.e. $E_{B_n}P_{n,B_n}^1 \hat{b}^k_{B_n}D^*_k(O)$ and $E_{B_n} P_{n,B_n}^1 \hat{b}^k_{1,B_n}D_k(O) \approx 0$.   This process ensures we have lowered the loss on our initial fit while solving the efficient influence curve equation, paving the way to compute an explicit second order remainder term telling us our bias and robustness properties of the estimator.  In this case, we are not doubly robust and the remainder term requires significant challenges.  Namely, our outcome model must be estimated such that the $L^2$ norm of the bias is $o_p(n^{-0.25})$ \parencite{blipvar}.  

\section{Donsker Condition}
In the original formulation of the CV-TMLE, we view the estimator as 10 plug-in estimators.  To compute each of the 10 estimators, the targeting step is performed on the validation set.  Since we can therefore condition on the training set from which the initial estimate is formed, we essentially have a fixed functions $\bar{Q}^0_{B_n}$ and $\hat{g}_{B_n}$, which we are fluctuating on the validation set with a one-dimensional parametric submodel.  Thus the entropy is very low for the class of functions containing $\bar{Q}^k_{B_n}$ in our above algorithm.  With our procedure the entropy is a little bigger in that the function, $\bar{Q}^k_{1,B_n}$, can be viewed as fixed, yet depending on an average over all validation sets (therefore very slightly inbred before the targeting step) as well as the fluctuation parameter, $\epsilon$, determined by the validation set.  The influence curve approximation, $D_{k, B_n}$, defined above, will thus have similarly low entropy as if we allowed another parameter in the parametric submodel.\\

Consider the following, which we pull out of Zheng and van der Laan, 2010, for the convenience of the reader.  

\begin{defn}
For a class of function, $\mathcal{F}$, whose elements are functions, $f$, that map observed data, $O$, to a real number, we define the entropy integral:
\[
Entro(\mathcal{F}) = \int_0^{\infty} \sqrt{\underset{Q}{\text{log sup}}N\left(\epsilon, \Vert F \Vert_{Q,2}, \mathcal{F}, L^2(Q) \right) d\epsilon}
\]
where $N\left(\epsilon, \mathcal{F}, L^2(Q) \right)$ is the covering number for $\mathcal{F}$, defined by the minimum number of balls of radius $\epsilon$ under the $L^2(Q)$ norm to cover $\mathcal{F}$.  $F$ is defined as the envelope of $\mathcal{F}$ or a function such that $\vert f \vert \leq F$ for all $f \in \mathcal{F}$.  
\end{defn}

Consider the following lemma (lemma 2.14.1 in ref van der Vaart and Wellner, 1996) \parencite{Vaart:1996aa}

\begin{lem}
Let $\mathcal{F}$ denote a class of measurable functions of $O$.  Let $G_n = \sqrt{n}(P_n - P_0)$.  Then 

\[
E(sup_{f\in \mathcal{F}} G_n f) \leq Entro(\mathcal{F}) \sqrt{P_0 F^2}
\]
\end{lem}

This lemma then yields the following results in Zheng and van der Laan, 2010.  Consider $\overset{\rightarrow k_n}{\epsilon_n}$, a sequence of $\epsilon_n^1,...,\epsilon^{k_0}_n$ that are the fluctuation parameters dependent on the draw from the data.  In the lemma below we assume the $k_0$ steps of a parametric fluctuation parameters converge in probability to a sequence of length $k_0$, a very weak assumption, the same as the estimated parameters of a parametric model converging to the truth in probability.  

\begin{lem}
Suppose $\Vert \overset{\rightarrow k_n}{\epsilon}- \overset{\rightarrow k_0}{\epsilon}\Vert \overset{P}{\rightarrow}0$.  For each sample split of $B_n$, we consider a class of measurable functions of $O$:
\[
\mathcal{F}\left( P_{n,B_n}^0 \right) = \left\{f_{\overset{\rightarrow}{\epsilon}} \left( P_{n,B_n}^0 \right)= f\left( \overset{\rightarrow}{\epsilon}, P_{n,B_n}^0 \right) - f\left( \overset{\rightarrow}{\epsilon_0} P_{n,B_n}^0 \right):\overset{\rightarrow}{\epsilon} \right\}
\]
where the index set contains $\epsilon_n$ with probability tending to 1.  For a deterministic sequence $\delta_n\rightarrow 0$, define subclasses

\[
\mathcal{F}_{\delta_n}\left( P_{n,B_n}^0 \right) = \left\{f_{\overset{\rightarrow}{\epsilon}} \in \mathcal{F}\left( P_{n,B_n}^0 \right) : \Vert \overset{\rightarrow}{\epsilon} - \overset{\rightarrow}{\epsilon_0} \Vert < \delta_n \right\}
\]
If for deterministic sequence $\delta_n\rightarrow 0 $ we have
\[
E\left\{ Entro(\mathcal{F}_{\delta_n}\left( P_{n,B_n}^0 \right)) \sqrt{P_0 F(\delta_n, P_{n,B_n}^0)^2} \right\} \rightarrow 0 \text{ as } n\rightarrow 0
\]
where $F(\delta_n, P_{n,B_n}^0)$ is the envelope of $\mathcal{F}_{\delta_n}\left( P_{n,B_n}^0 \right)$, then
\[
\sqrt{n}(P_{n,B_n}^1 - P_0) \left\{f(\overset{\rightarrow}{\epsilon_n}, P_{n,B_n}^0) - f(\overset{\rightarrow}{\epsilon_0}, P_0) \right\} = o_P(1)
\]
\end{lem}

We note to the reader that we keep lemma 3.2 identical to what was in Zheng and van der Laan, 2010, except we do not condition solely on $P_{n,B_n}^0$ when defining $\mathcal{F}\left( P_{n,B_n}^0 \right)$.  Such does not at all affect the truth of the lemma.

\subsection{Remainder Term}

Our estimate minus the truth is, using notation in Zheng and van der Laan, 2010, where $\overset{\rightarrow}{k_n},$ indicates the $k_n$ iteration, we have

\[
\Psi^{k_n}(P_n) = E_{B_n} \hat{\Psi}_{B_n} (P_{n})
\]

The second order remainder, $R_2(\cdot)$, can be written:

\begin{eqnarray*}
\Psi^{k_n}(P_n) - \Psi(P_0) &=& E_{B_n}(P_{n,B_n}^1 - P_0)D_{\overset{\rightarrow}{k_n}, B_n} + R_2(P_n, P_0)\\
&=& -E_{B_n}P_0 D_{\overset{\rightarrow}{k_n}, B_n} + R_2(P_n, P_0)
\end{eqnarray*}

Assuming the remainder is $o_P(1/\sqrt{n})$, we then get that  
\[
\Psi^{k_n}(P_n) - \Psi(P_0) = E_{B_n}(P_{n,B_n}^1 - P_0 )D_{\overset{\rightarrow}{k_n}, B_n} + o_P(1\sqrt{n}) =-E_{B_n}P_0 D_{\overset{\rightarrow}{k_n}, B_n} + R_2(P_n, P_0)
\]

since our procedure solves $E_{B_n} P_{n,B_n}^1 D_{\overset{\rightarrow}{k_n}, B_n} = 0$.  As discussed, we can quite easily satisfy lemma 3.2 for the function class containing $D_{\overset{\rightarrow}{k_n}, B_n}$.  Again, assuming the remainder is $o_P(1/\sqrt{n})$ our estimator is asymptotically efficient if $D_{\overset{\rightarrow}{k_n}, B_n}$ converges to the true influence curve in $L^2(P_0)$ \parencite{Laan:2006aa}.  For variance of the treatment effect, the remainder term conditions are no more strict than for the original formulation of the CV-TMLE. 

\section{Conclusion}
This slight adjustment to the CV-TMLE algorithm is easier to implement and retains the same theoretical properties as in our example here.  It remains to be more formally generalized to include a class of TMLE's for which it is valid but the author feels the example used here-in gives the reader sufficient intuition to understand when such can be done.  For one, it is obvious if any polynomial factor of a mean (assuming the mean converges) appears as a factor in the clever covariate, then the entropy will be similarly small, so this procedure covers many examples one might find in practice.  The procedure overlaps exactly with the originally formulated CV-TMLE with many common parameters where the clever covariates contain no empirical means.  It is a subject for future research whether this procedure has any advantages in finite samples.  Simulations have shown no appreciable difference in performance for VTE.      

\newpage
\printbibliography
\end{document}